# Two Dimensional Phosphorus Oxides as Energy and Information Materials

Wei Luo[1,2] and Hongjun Xiang[1,2]*

[1]*Key Laboratory of Computational Physical Sciences (Ministry of Education), State Key Laboratory of Surface Physics, and Department of Physics, Fudan University, Shanghai 200433, P. R. China*

[2]*Collaborative Innovation Center of Advanced Microstructures, Nanjing 210093, P. R. China*

Email: hxiang@fudan.edu.cn

**Abstract:** Two-dimensional (2D) black phosphorus (i.e., phosphorene) has become a rising star in electronics. Recently, 2D phosphorus oxides with higher stability have been synthesized. In this work, we systematically explore the structures and properties of 2D phosphorus oxides on the basis of global optimization approach and first-principles calculations. We find that the structural features of 2D phosphorus oxides $P_xO_y$ vary with the oxygen concentration. When the oxygen content is poor, the most stable 2D $P_xO_y$ can be obtained by adsorbing O atoms on phosphorene. However, when the oxygen concentration becomes rich, stable structures are no longer based on phosphorene and will contain P-O-P motifs. For the 2D $P_4O_4$, we find that it has a direct band gap (about 2.24 eV), good optical absorption, and high stability in water, suggesting that it may be good candidate for photochemical water splitting application. Interestingly, 2D $P_2O_3$ adopt two possible stable ferroelectric structures ($P_2O_3$-I and $P_2O_3$-II) as the lowest energy configurations within a given layer thickness. The electric polarizations of $P_2O_3$-I and $P_2O_3$-II are perpendicular and parallel to the lateral plane, respectively. We propose that 2D $P_2O_3$ could be used in a novel nanoscale multiple-state memory device. Our work suggests that 2D phosphorus oxides may be excellent functional materials.

In recent years, two dimensional (2D) materials have attracted lots of interests for their novel physical properties and potential applications[1]. Graphene is a representative 2D material[1a-d] in which many exotic phenomena were discovered due to the presence of a Dirac-type band dispersion. At the same time, the absence of a band gap unfortunately limits its applications in electronic and optoelectronic devices. Subsequently, monolayer transition metal dichalcogenides (TMDs) were studied extensively due to their intrinsic direct band gaps[1e-g]. Despite multiple researches report that the electronic and optoelectronic devices based on monolayer TMDs have shown high on/off ratio and high responsivity, the carrier mobility of these TMDs members is still much lower than graphene[2]. Most recently, 2D monolayer black phosphorus (BP), i.e., phosphorene, has attracted much attention since it not only has a direct band gap of about 2.0 eV[3], but also high carrier mobility[4]. Interestingly, other phosphorene-like materials were predicted to display high carrier mobilities and a broad range of band gaps[5]. These unique properties make phosphorene superior to graphene in electronic and optoelectronic applications. In fact, it was demonstrated that the field-effect transistor (FET) made of few-layer black phosphorus presented high on/off ratio[4a]. Experiment also showed that photodetector made of multilayer black phosphorus can exhibit high-contrast images both in the visible as well as in the infrared spectral regime[6]. Due to the inherent orthorhombic waved structure of phosphorene, the carrier mobility is highly anisotropic[7] and its electronic properties can be easily tuned by strain, which could be useful in some special applications[8].

Although phosphorene is a promising material with many novel properties, it has a well-known drawback. It degrades easily in the oxygen and humidity environment[9]. Therefore, it is important to understand the oxidation mechanism in phosphorene. This is difficult to be investigated directly through experiments since oxidation may lead to amorphous structures that cannot be probed by diffraction, and since most techniques that measure the real space structure lead to further sample degradation. Theoretically, Ziletti *et al.* found a single O atom tend to adsorb on one phosphorus atom, forming the dangling-type structure[9a]. This is because the electron lone pair of the sp$^3$ hybridized P atom could interact with an oxygen atom which needs two more electrons to satisfy the octet rule. Thus, the interaction mechanism between phosphorene and a few oxygen molecules and oxygen atoms is now clear[10]. But, the structure and properties of recently synthesized layered phosphorus oxide compounds ($P_xO_y$) with a high oxygen concentration are not well understood[11]. Although some model structures for 2D phosphorus oxides have been built manually[12], it is possible that these model structures are not necessarily stable. On the other hand, like graphene oxides[13], 2D phosphorus oxides themselves may serve as novel functional materials with fascinating properties. In fact, Lu *et al.* experimentally demonstrated that phosphorus oxides and suboxides not only have tunable band gaps, but also are much more stable compared with pristine phosphorene under ambient condition. They further showed that these 2D phosphorus oxides could be used in multicolored display and toxic gas sensor applications[14]. So, a detailed understanding of the structure of 2D phosphorus oxides will not only provide deeper insight into the degradation mechanism, but also may lead to the discovery of new functional 2D materials.

In this work, we systematically predict the lowest energy structures of 2D phosphorus oxides with different oxygen concentrations through our newly developed global optimization approach[15]. We find that 2D phosphorus oxides keep the phosphorene framework with dangling P=O motifs when the oxygen concentration is low, while there will exist P-O-P motifs when the oxygen concentration is higher than one third. In the most stable structures of $P_4O_4$ and $P_2O_3$ with the thickness less than 3.2 Å, there are only P-O-P motifs but no dangling P=O motif. Interestingly, the most stable structure of 2D $P_4O_4$, i.e., $P_4O_4$-I, has a direct band gap of 2.24 eV, good optical absorption, appropriate band edge positions, and high stability in water, suggesting that it may be used for photocatalytic water splitting. For 2D $P_2O_3$, there are two stable ferroelectric structures (namely $P_2O_3$-I and $P_2O_3$-II) with non-zero out-of-plane and in-plane polarization, respectively, suggesting that 2D $P_2O_3$ may be used in novel multiple-state memory devices. Our work shows that 2D phosphorus oxides could act as novel functional materials.

We search the stable structures of 2D phosphorus oxides with different oxygen concentrations through a global optimization approach. For some oxygen concentrations, we find new structures with total energies lower than the previously proposed structures. In the following, we will first discuss the structure features of 2D phosphorus oxides revealed from our particle swarm optimization (PSO) simulations. Then, we will focus on 2D $P_4O_4$ and $P_2O_3$ to demonstrate that they are promising materials for the use in photochemical water splitting and ferroelectric multiple-state memory, respectively.

From our PSO simulations, we find that the stable structures of phosphorus oxides display different motifs when oxygen concentration varies. This is because the Coulomb interactions between oxygen ions will destabilize the dangling-BP type structure in the high oxygen concentration case. To be more specific, for $P_8O_1$ and $P_6O_1$, their lowest-energy structures are of the dangling-BP type, as shown in Figure 1. For $P_4O_1$, its 2D structure also belongs to the dangling-BP type which was first proposed by Wang et al.[12a] (see Figure 1). But surprisingly and interestingly, we find that $P_4O_1$ has another special one dimensional (1D) tubular structure (supporting information, Figure S1). This 1D structure has lower total energy (about 14 meV/atom) than its 2D configuration. Note that the 1D $P_4O_1$ with a 3.3 Å diameter may be the smallest nanotube since the smallest experimentally observed carbon nanotube has a diameter of 4 Å[16]. For $P_2O_1$, it contains not only the dangling P=O motifs but also the P-O-P motifs (hereafter referred as $P_2O_1$-I, see Figure 1). Its total energy is lower than that of previous proposed $P_2O_1$ structure (referred as $P_2O_1$-IV, see supporting information, Figure S4) [12a] by about 174 meV/atom despite of the fact that the $P_2O_1$-I structure is thinner than the previous $P_2O_1$ structure by about 1.0 Å. It is worth stressing that $P_2O_1$-I is unusual and complicated. There exist four-member rings and eight-member rings formed by phosphorus atoms. Since its framework is completely different from that of phosphorene, it is almost impossible to build it manually, indicating that our global optimization approach is powerful. The most stable structures of $P_4O_4$ and $P_2O_3$ are shown in Figure 2(a) and Figure 5(b), respectively. We can find that both of them belong to the bridge-type structure without the P=O motifs any more. Previously, it was suggested that the band gap of phosphorus oxides increases monotonously with the oxygen concentration[12b, 14]. Here, we find that this is not necessarily true: The band gap of our $P_2O_1$ structure is about 2.85 eV, which is larger than that (2.24 eV) of the stable structure of $P_4O_4$ (i.e. $P_4O_4$-I). This suggests that 2D phosphorus oxides have rich structural and electronic properties. As expected, the average binding energy per phosphorus atom increases with the oxygen concentration (supporting information, Figure S2).

The lowest energy structure of $P_4O_4$ (i.e., $P_4O_4$-I) within the 3.2 Å thickness only contains P-O-P bridge motifs, as shown in Figure 2(a). Six phosphorus atoms and four oxygen atoms form a ring. There are two P-P dimers in a primitive cell. Each oxygen atom is two-fold coordinated, and each phosphorus atom bonds with one phosphorus atom and two oxygen atoms. The phosphorus atom takes the $sp^3$ hybridization with an electron lone pair, while there are two electron lone pairs for each $sp^3$ hybridized oxygen atom. This chemical bonding analysis suggests that $P_4O_4$-I is a semiconductor. The calculated band structure indeed confirms the semiconducting nature. As can be seen from Figure 3(a), $P_4O_4$-I has a direct band gap of 2.24 eV at $\Gamma$ point. The reason why the valence band maximum (VBM) located at $\Gamma$ point is because of the interaction between the P-P σ bonding states (see supporting information for a detailed analysis). Moreover, we find the band dispersion near $\Gamma$ is large, indicating a high mobility. The electron effective mass with the local density approximation (LDA) functional is computed to be 0.58 $m_0$, which is even smaller than that (0.68 $m_0$) of phosphorene. The optical absorption spectrum calculated with the HSE functional is plotted in Figure 4(b). From it, we can see the absorption starting from 2.24 eV, indicating the dipole transition between conduction band minimum (CBM) and VBM is allowed. This can be understood within the group theory. The point group of $P_4O_4$-I is $C_{2h}$, which has four one dimensional irreducible representations. We find that the wave function of CBM and VBM belong to odd $A_u$ and even $B_g$ representations, respectively. This explains the dipole-transition between the band edge states.

As we mentioned above, the band gap of $P_4O_4$-I is even smaller than $P_2O_1$-I. In order to understand this unusual phenomenon, we plot wave functions of VBM and CBM states of $P_4O_4$-I in Figure 3(c) and (d). We find that the VBM state is mainly contributed by the P-P σ bonding states. While for the CBM state, the wavefunction is mainly distributed between one phosphorus atom and its next-nearest neighbor in a six phosphorus ring (namely, $P_1$ and $P_2$ atom). The interaction between $P_1$ and $P_2$ atoms is very important to the low energy of the CBM state, and thus is responsible for the small band gap of $P_4O_4$-I. The uncommon long-range P-P interaction is due to the twisting of $P_6O_4$ rings in $P_4O_4$-I (For more detailed discussions, see supporting information).

Our above results indicate that $P_4O_4$-I may be appropriate for photoelectrochemical (PEC) water splitting application[17]. As we know, in order to be a good photochemical water splitting material, the positions of band edges should be suitable for solar-driven water splitting[18]. The estimated band edge positions with respect to the vacuum level are shown in Figure 4(a). The hydrogen evolution potential and oxygen evolution potential are marked with black dot and green dot respectively. It can be seen that the CBM of $P_4O_4$-I is higher than the hydrogen evolution potential by about 0.67 eV and the VBM of $P_4O_4$-I is below the oxygen evolution potential by about 0.37 eV. These band edge positions are suitable for PEC water splitting. We investigate the interaction between water and $P_4O_4$-I with first-principles molecular dynamics (MD) simulation (supporting information, Figure S3). Our results show that $P_4O_4$-I can be stable in water at room temperature. In addition, the adsorption energies of a water molecule on $P_4O_4$-I and phosphorene are found to be -256 meV and -283 meV, respectively. This suggests that $P_4O_4$-I is more stable than phosphorene in water[19].

Three dimensional ferroelectric materials are widely explored both experimentally and theoretically[20]. In two dimension, the depolarizing field was believed to suppress the ferroelectric dipoles perpendicular to the film surface, leading to the disappearing of ferroelectricity with thickness less than a certain value[21]. For this reason, only a few works focused on 2D ferroelectrics in the past[21b, 22]. Recently, the 1T monolayer $MoS_2$ and 2D honeycomb binary compounds were predicted to be 2D ferroelectric materials. But their ferroelectric properties still await experimental confirmation. So, discovering new 2D ferroelectric materials will not only help us to understand new physical mechanism for 2D ferroelectricity, but also accelerate the applications of 2D ferroelectrics.

From our PSO simulation, we find two kinds of ferroelectric structures for $P_2O_3$

(namely, $P_2O_3$-I and $P_2O_3$-II), as shown in Figure 5(a) and (b). For $P_2O_3$-I, it is the lowest energy structure with thickness less than 1.4 Å. The phosphorus atoms form a honeycomb lattice and each phosphorus atom is surrounded by three oxygen atoms. Thus, there are only P-O σ bonds, resulting in a large band gap (about 5.79 eV). All phosphorus atoms are located in the top-plane and all oxygen atoms are located in the bottom-plane. Hence, $P_2O_3$-I has a non-zero electric polarization perpendicular to the lateral plane. The presence of a perpendicular ferroelectric polarization in $P_2O_3$-I can be explained as following. Although the pure electrostatic interaction between the $P^{3+}$ ion and $O^{2-}$ ion favors a flat structure, the instability in flat-$P_2O_3$ is due to the presence of lone-pair electrons of the $P^{3+}$ ion. As shown in Figure 6, the interaction between P $3p_z$ and 3s orbitals in flat-$P_2O_3$ is forbidden by symmetry. In contrast, P $3p_z$ orbital can mix with P 3s orbital to lower the energy level in $P_2O_3$-I. Since this level will be occupied by the lone pair electrons, the total energy becomes lower than that of flat-$P_2O_3$. This is rather similar to the mechanism of the buckling of the $NH_3$ molecule. The ferroelectric mechanism in $P_2O_3$-I is different from that in hexagonal ABC hyperferroelectric[23] where small effective charges and large dielectric constants play a role. According to Garrity et al.[23], hyperferroelectrics refers to a class of proper ferroelectrics which polarize even when the depolarization field is unscreened. In this sense, $P_2O_3$-I is the thinnest hyperferroelectrics originated from a new lone-pair mechanism.

$P_2O_3$-II is the lowest energy structure of $P_2O_3$ with the thickness less than 3.2 Å. Its topology is similar with that of $P_2O_3$-I. But unlike $P_2O_3$-I, the phosphorus atoms are no longer in the same plane and so do oxygen atoms, leading to zero polarization perpendicular to the lateral plane. But it has a non-zero in-plane polarization due to the collective oxygen displacements along the $y$ axis [see Figure 5(b)]. The total energy of $P_2O_3$-II is lower than $P_2O_3$-I by about 34 meV/atom. However, we note that $P_2O_3$-II has a larger thickness than $P_2O_3$-I (1.46 Å vs. 0.80 Å). To estimate the switching barrier and magnitude of electric polarization, we consider a paraelectric phase of $P_2O_3$ [$P_2O_3$-III in Fig 5(c)] in which the oxygen plane is sandwiched between the two phosphorus planes. The paraelectric phase $P_2O_3$-III is a semiconductor with a LDA band gap of

3.41 eV. The energy barrier between ferroelectric $P_2O_3$-I ($P_2O_3$-II) and paraelectric phase is about 75 meV/atom (109 meV/atom). So, it is possible to switch the ferroelectric phase to the paraelectric phase through applying the external electric field. The electric polarizations of $P_2O_3$-I and $P_2O_3$-II are found to be $2.4\times10^{-12}$ C/m and $1.2\times10^{-12}$ C/m, respectively.

For $P_2O_3$-I, the electric polarization can be either along $z$ or $-z$ directions. For $P_2O_3$-II, there are six possible ferroelectric states with in-plane electric polarizations because of the six-fold symmetry of the paraelectric state. Therefore, totally there are eight different ferroelectric states for 2D $P_2O_3$. With the electric field, we can change the direction of electric polarization and the nanoscale multiple-state (8-state) memory device can be possibly realized (see Figure 7).

We note that these 2D phosphorus oxides are thermally and kinetically stable (supporting information, Figure S6). Finally, we proposed two possible ways to synthesize these materials. One way to obtain $P_4O_4$ and $P_2O_3$ is to partially oxidize the phosphorene by ozone or oxygen plasma with controlled oxygen concentration. Another way is to reduce the phosphorus oxides with high oxygen concentration by the chemical reduction method, which was successfully adopted to obtain partially oxidized graphene[24].

In conclusion, we systematically predict the lowest energy structures of 2D $P_8O_1$, $P_6O_1$, $P_4O_1$, $P_2O_1$, $P_4O_4$ and $P_2O_3$ with the global optimization method. We find that the features of stable structures vary with oxygen concentration. If the oxygen concentration is low, 2D $P_xO_y$ structures are based on phosphorene with dangling P=O motifs, as reported by Zeilt *et al.* With the increase of oxygen concentration, 2D $P_xO_y$ structures will most likely exhibit the P-O-P motifs. We further show that 2D $P_xO_y$ may have unique properties for functional materials. $P_4O_4$-I may be good candidate for PEC water splitting application since it has an appropriate band gap, good optical absorption, and high stability in water. Both $P_2O_3$-I and $P_2O_3$-II are 2D ferroelectrics. In particular, $P_2O_3$-I may be the thinnest hyperferroelectrics originated from a new mechanism. We propose that 2D $P_2O_3$ could be used in nanoscale multiple-state memory devices in the future.

**Experimental Section**

In this work, density functional theory (DFT) method is used for structural relaxation and electronic structure calculation. The ion-electron interaction is treated by the projector augmented-wave (PAW)[25] technique as implemented in the Vienna ab initio simulation package[26]. The exchange-correlation potential is treated by LDA[27]. For structural relaxation, all the atoms are allowed to relax until atomic forces are smaller than 0.01 eV/Å. The 2D k-mesh is generated by the Monkhorst-Pack scheme. To avoid the interaction between neighboring layers, the vacuum thickness is chosen to be 12 Å. To obtain more reliable results for electronic and optical properties, the HSE06 functional[28] is adopted since LDA underestimates the band gap. The band edge positions are estimated by aligning the vacuum level with respect to the electrostatic potential in the vacuum region of the supercell.

In our implementation, for each 2D structure, we first randomly select a layer group[15] instead of a planar space groups[29]. The lateral lattice parameters and atomic positions are then randomly generated but confined within the chosen layer group symmetry. Subsequently, local optimization including the atomic coordinates and lateral lattice parameters is performed for each of the initial structures. In the next generation, a certain number of new structures (the best 60% of the population size) are generated by PSO[30]. The other structures are generated randomly, which is critical to increase the structure diversity. When we obtain the new 2D structures by the PSO operation or random generation, we make sure that the thickness of the 2D structure is smaller than the given thickness.

Extensive PSO simulations are performed to find out the global stable structures of 2D $P_8O_1$, $P_6O_1$, $P_4O_1$, $P_2O_1$, $P_4O_4$, and $P_2O_3$. We set the population size to 30 and the number of generations to 20. The total number of atoms in the unit cell is less than 16. We consider five different thicknesses (between 1 and 4 Å) for each system. In addition, we repeat twice of each calculation in order to make results reliable.

**Keywords:** two dimensional, phosphorus oxides, photochemical water splitting, ferroelectricity, density functional theory, global structure optimization.**Associated Content**

**Supporting Information**

Supporting Information accompanies this paper at http://onlinelibrary.wiley.com. Correspondence and requests for materials should be addressed to H.X.

**Author Information**

Corresponding Author

*E-mail: hxiang@fudan.edu.cn (H. J. Xiang).

**Acknowledgements**

Work was supported by NSFC, the Special Funds for Major State Basic Research, Research Program of Shanghai Municipality and MOE, Program for Professor of Special Appointment (Eastern Scholar), Qing Nian Ba Jian Program, and Fok Ying Tung Education Foundation.

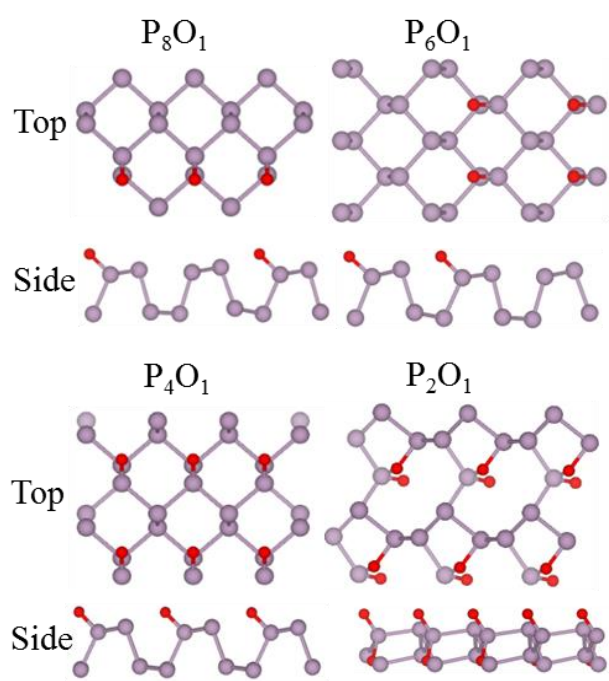

**Figure 1.** The lowest energy structures of $P_8O_1$, $P_6O_1$, $P_4O_1$, and $P_2O_1$ with thickness less than 3.2 Å. For $P_8O_1$, $P_6O_1$, and $P_4O_1$, they are still based on the phosphorene structure. But for $P_2O_1$ (i.e., $P_2O_1$-I), it is no longer based on phosphorene structure and it has P-O-P motif besides the P=O motifs.

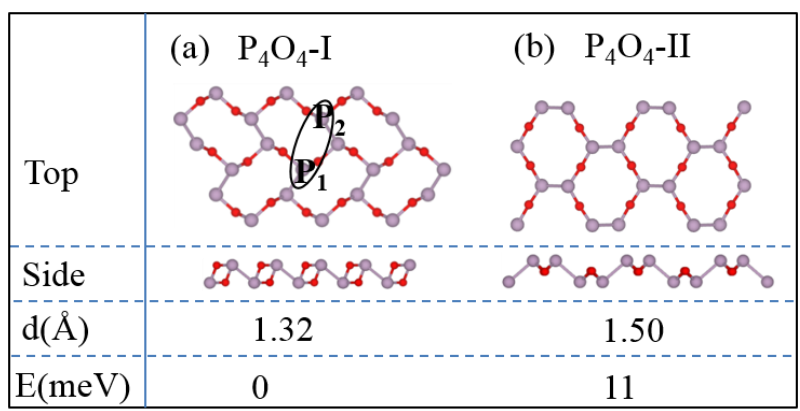

**Figure 2.** a) The lowest energy structure (i.e., $P_4O_4$-I) of $P_4O_4$ with thickness less than 3.2 Å. The $P_1$ and $P_2$ atoms are denoted by a black ellipse. b) Another structure (i.e., $P_4O_4$-II) of $P_4O_4$. $P_4O_4$-II has a similar topology as $P_4O_4$-I, but no twisting of $P_6O_4$ rings. $P_4O_4$-II has a higher energy by 11 meV/atom than $P_4O_4$-I. "d" is the thickness of 2D structures.

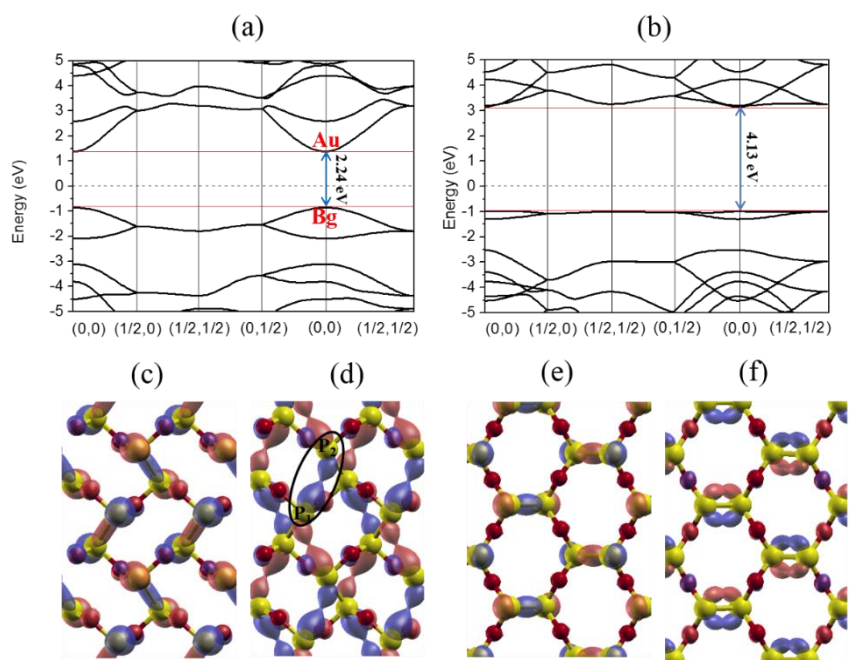

**Figure 3.** a) Band structure of $P_4O_4$-I from the HSE06 calculation. It has a direct band gap of about 2.24 eV at Γ point. b) Band structure of $P_4O_4$-II from the HSE06 calculation. c) The VBM wave function of $P_4O_4$-I, which is contributed by the P-P bonding state. d) The CBM wave function of $P_4O_4$-I, which mainly distribute around $P_1$ and $P_2$ atoms. e) The VBM wave function of $P_4O_4$-II. Similar with $P_4O_4$-I, it is also contributed by the P-P bonding state. f) The CBM wave function of $P_4O_4$-II. Unlike

$P_4O_4$-I, it is mainly contributed by π bonds of the P-P dimer.

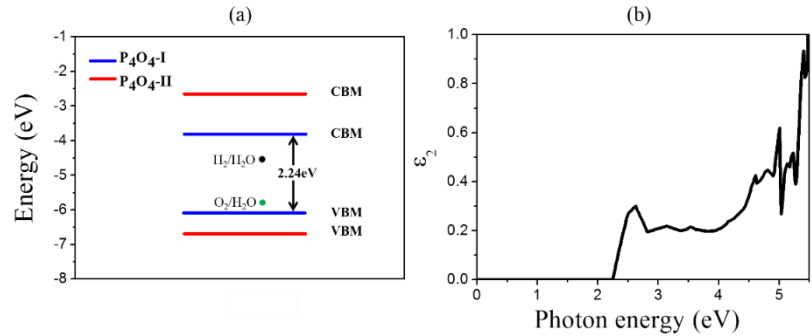

**Figure 4.** a) Band edges of $P_4O_4$-I and $P_4O_4$-II calculated with the HSE06 functional, referenced to the vacuum level. The green and black dots represent the oxygen and hydrogen evolution potentials in the case of pH = 0, respectively. b) Imaginary part of the dielectric function of $P_4O_4$-I from the HSE06 calculation. It suggests that the optical absorption increase rapidly above the band gap.

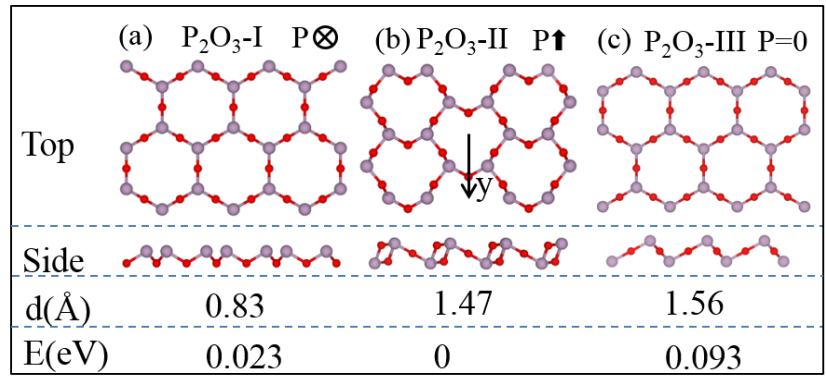

**Figure 5.** a) Top and side views of $P_2O_3$-I, which is the lowest energy $P_2O_3$ with thickness less than 1.4 Å. Each phosphorus atom is surrounded by three oxygen atoms. $P_2O_3$-I is ferroelectric with an out-of-plane electric polarization. b) Top and side views of $P_2O_3$-II, which is the lowest energy $P_2O_3$ with thickness less than 3.2 Å. $P_2O_3$-I is ferroelectric with an in-plane electric polarization. c) Top and side views of the paraelectric $P_2O_3$ structure (i.e., $P_2O_3$-III). Oxygen atoms are inversion centers.

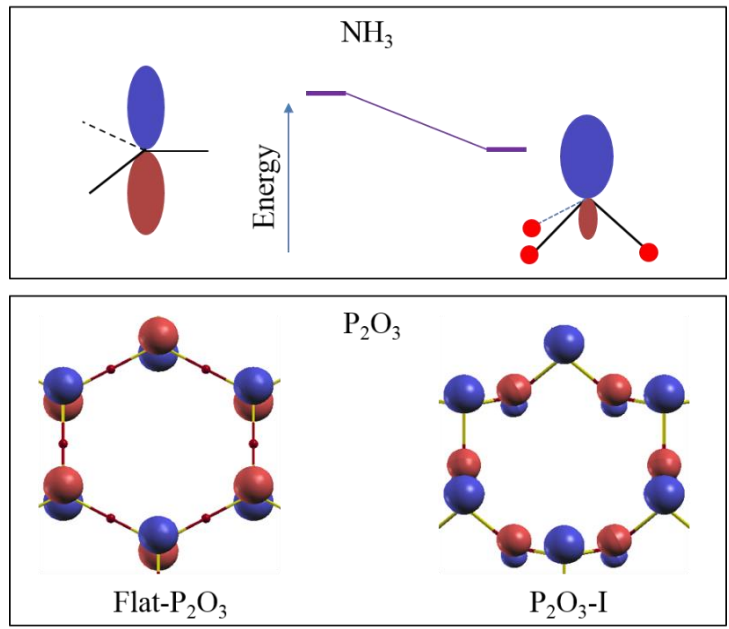

**Figure 6.** The schematic diagrams of highest occupied molecule orbitals of $NH_3$ and $P_2O_3$. For the flat configuration, the $p_z$ orbitals of phosphorus atoms can not hybridize with the s orbitals. While for the buckled configuration, the $p_z$ orbitals can mix with s orbitals and lower the total energy.

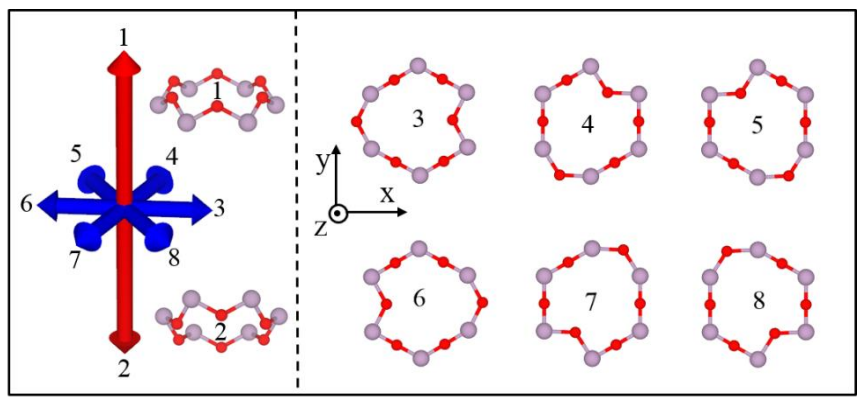

**Figure 7.** Schematic illustration of the concept of multiple state memory based on 2D $P_2O_3$. In total, there are eight possible ferroelectric states in 2D $P_2O_3$.